\documentclass[preprintnumbers,showpacs,amsmath,amssymb,floatfix,twocolumn,prd,superscriptaddress,nofootinbib]{revtex4}
\usepackage{graphicx}
\usepackage{epsfig}
\usepackage{bm}
\usepackage{amsfonts}

\newcommand{\la}{\lambda}

\newcommand{\bear}{\begin{eqnarray}}

\newcommand{\eear}{\end{eqnarray}}

\begin{document}

\title{The generalized second law of thermodynamics in Ho\v{r}ava-Lifshitz cosmology}

\author{Mubasher Jamil}
\email{mjamil@camp.nust.edu.pk} \affiliation{Center for Advanced
Mathematics and Physics, National University of Sciences and
Technology,\\ Rawalpindi, 46000, Pakistan}

\author{Emmanuel N. Saridakis}
\email{msaridak@phys.uoa.gr} \affiliation{Department of Physics,
University of Athens, GR-15771 Athens, Greece}

\author{M. R. Setare}
\email{rezakord@ipm.ir} \affiliation{Department of Science of
Bijar, University of Kurdistan, Bijar, Iran }

\begin{abstract}
We investigate the validity of the generalized second law of
thermodynamics in a universe governed by Ho\v{r}ava-Lifshitz
gravity. Under the equilibrium assumption, that is in the
late-time cosmological regime, we calculate separately the entropy
time-variation for the matter fluid and, using the modified
entropy relation, that of the apparent horizon itself. We find
that under detailed balance the generalized second law is
generally valid for flat and closed geometry and it is
conditionally valid for an open universe, while beyond detailed
balance it is only conditionally valid for all curvatures.
Furthermore, we also follow the effective approach showing that it
can lead to misleading results. The non-complete validity of the
generalized second law could either provide a suggestion for its
different application, or act as an additional problematic feature
of Ho\v{r}ava-Lifshitz gravity.
\end{abstract}


\maketitle

\section{Introduction}

Almost one year ago, Ho\v{r}ava proposed a power-counting
renormalizable, ultra-violet (UV) complete theory of gravity
\cite{hor2,hor1,hor3,hor4}. Although presenting an infrared (IR)
fixed point, namely General Relativity, in the  UV the theory
possesses a fixed point with an anisotropic, Lifshitz scaling
between time and space. These novel features led many authors to
examine and extend the properties of the theory itself
\cite{Volovik:2009av,Cai:2009ar,Cai:2009dx,Orlando:2009en,Nishioka:2009iq,Konoplya:2009ig,Charmousis:2009tc,Li:2009bg,Nojiri:2009th,Kluson:2009rk,Sotiriou:2009gy,
Sotiriou:2009bx,Germani:2009yt,Chen:2009bu,Chen:2009ka,Shu:2009gc,Bogdanos:2009uj,Afshordi:2009tt,Myung:2009ur,Alexandre:2009sy,
Blas:2009qj,Capasso:2009fh,Chen:2009vu,Koyama:2009hc,Papazoglou:2009fj,Kluson:2009xx,Kluson:2010aw},
and furthermore to apply it as a cosmological framework,
constructing the so-called Ho\v{r}ava-Lifshitz cosmology
\cite{Calcagni:2009ar,Kiritsis:2009sh}. Amongst the very
interesting physical implications are the novel
 solution subclasses
\cite{Lu:2009em,Nastase:2009nk,Colgain:2009fe,Ghodsi:2009rv,Minamitsuji:2009ii,Ghodsi:2009zi,Wu:2009ah,
Cho:2009fc,Boehmer:2009yz,Momeni:2009au,Setare:2009sw,Kiritsis:2009vz,Cai:2010ud,
Chaichian:2010yi,Carloni:2009jc,Leon:2009rc,Myung:2009if,Bakas:2009ku,Myung:2010qg},
the gravitational wave production
\cite{Mukohyama:2009zs,Takahashi:2009wc,Koh:2009cy,Park:2009gf,Park:2009hg,Myung:2009ug},
the perturbation spectrum
\cite{Mukohyama:2009gg,Piao:2009ax,Gao:2009bx,Chen:2009jr,Gao:2009ht,Cai:2009hc,
Wang:2009yz,Kobayashi:2009hh,Wang:2009azb,Kobayashi:2010eh},
the matter bounce
\cite{Brandenberger:2009yt,Brandenberger:2009ic,Cai:2009in,Suyama:2009vy,Czuchry:2009hz,Gao:2009wn},
 the dark energy phenomenology
\cite{Saridakis:2009bv,Park:2009zr,Appignani:2009dy,Setare:2009vm,Garattini:2009ns,Setare:2010wt,Jamil:2010vr},
 the astrophysical phenomenology
\cite{Kim:2009dq,Iorio:2009qx,Iorio:2009ek,Izumi:2009ry,Harko:2009qr},
and the observational constraints on the theory
\cite{Dutta:2009jn,Dutta:2010jh,Ali:2010sv}.

A specific direction of the research on the topic is the
investigation of the thermodynamic properties of
Ho\v{r}ava-Lifshitz gravity, a project that is crucially connected
to the black hole properties in such a theory
\cite{Danielsson:2009gi,Cai:2009pe,Myung:2009dc,Kehagias:2009is,Myung:2009va,
Cai:2009qs,Mann:2009yx,Bertoldi:2009vn,Park:2009zra,Castillo:2009ci,
BottaCantcheff:2009mp,Lee:2009rm,Varghese:2009xm,Cai:2009ph,
Kiritsis:2009rx,Majhi:2009xh,Greenwald:2009kp,Setare:2010zd,
Setare:2010ic,Lobo:2010hv,Setare:2010gi,Wei:2010yw}.
In particular, due to the well known connection between
thermodynamics and gravity (see \cite{Padmanabhan:2009vy} and
references therein), one is able to express the field equations as
a first law of thermodynamics (however the inverse procedure is
not always possible, that is starting from thermodynamics to
extract the general field equations, without this implying that
the specific gravitational theory is inconsistent \footnote{We
thank R. G. Cai for this comment.}). Additionally, the
aforementioned thermodynamic interpretation of the field equations
can be extended in cosmology, and it is applicable to any horizon
provided that the gravitational theory is diffeomorphism invariant
 \cite{Padmanabhan:2009ry}.
Having in mind that Ho\v{r}ava-Lifshitz gravity is not
diffeomorphism invariant, and thus in cosmological frameworks one
has to use the apparent horizon \cite{Bak:1999hd} instead of an
arbitrary one, some authors have investigated the connection
between the first law of thermodynamics and the Friedmann
equations \cite{Wang:2009rw,Cao:2010xx,Shu:2010nv,Wei:2010ww}.

In order to interpret the Friedmann equations of
Ho\v{r}ava-Lifshitz cosmology  thermodynamically, one faces two
possible approaches. The first and correct one is to consider a
universe containing only the matter fluid, and calculate the
horizon entropy using the modified relation for the black hole
entropy in Ho\v{r}ava-Lifshitz gravity \cite{Wei:2010ww}.
 The second approach is to absorb
all the extra information of
 Ho\v{r}ava-Lifshitz cosmology in an effective dark energy
 component, and thus to consider a universe containing the matter
 and the effective dark energy fluid in a general-relativity
 background, that is calculate the horizon
entropy using the standard relation for the black hole entropy
\cite{Wang:2009rw,Cao:2010xx}. Interestingly, while with the
second approach one obtains exactly the standard Friedmann
equations, with the first one he acquires some  additional
corrections, and the two results coincide only in the
zero-curvature or in the low energy limit  \cite{Wei:2010ww} (see
also \cite{Zhang:2010hi}). However, we should stress that the
first approach is the robust one since it treats the gravitational
sector separately and completely, taking into account its radical
effects on the geometry, while the second approach defines naively
an effective ``gravitationally-originated'' fluid and treating it
as a conventional fluid, not taking into account that gravity is
not only an ``actor'' but a ``director'', too. Finally, the better
theoretical background of the first approach becomes obvious in
the fact that it is followed in all the thermodynamic studies of
various alternative gravitational theories
\cite{Paranjape:2006ca}.

In the present work we are interested in investigating the
validity of the  generalized second law of thermodynamics in the
context of Ho\v{r}ava-Lifshitz cosmology. We follow the exact and
robust approach, that is we use the modified entropy relation as
it has been calculated in the specific context of
Ho\v{r}ava-Lifshitz gravity. Under the equilibrium assumption
between the universe interior and the horizon, which is expected
to ba valid at late cosmological times, we find that the
generalized second law is only conditionally valid. For
completeness, we also follow the discussed effective approach,
showing that it can lead to misleading results. The plan of the
work is as follows: in section \ref{model} we present the
cosmology of a universe governed by Ho\v{r}ava-Lifshitz gravity
and in section \ref{GSL} we investigate the validity of the
generalized second law of thermodynamics. In section \ref{GSLeff}
we perform the same analysis following the effective approach.
Finally, in section \ref{conclusions} we discuss and we summarize
the obtained results.

\section{Ho\v{r}ava-Lifshitz cosmology}
\label{model}

In this section we briefly review the scenario where the
cosmological evolution is governed by Ho\v{r}ava-Lifshitz gravity
\cite{Calcagni:2009ar,Kiritsis:2009sh}. The dynamical variables
are the lapse and shift functions, $N$ and $N_i$ respectively, and
the spatial metric $g_{ij}$ (roman letters indicate spatial
indices). In terms of these fields the full metric is written as:
\begin{eqnarray}
ds^2 = - N^2 dt^2 + g_{ij} (dx^i + N^i dt ) ( dx^j + N^j dt ) ,
\end{eqnarray}
where indices are raised and lowered using $g_{ij}$. The scaling
transformation of the coordinates reads: $
 t \rightarrow l^3 t~~~{\rm and}\ \ x^i \rightarrow l x^i
$.

The gravitational action is decomposed into a kinetic and a
potential part as $S_g = \int dt d^3x \sqrt{g} N ({\cal L}_K+{\cal
L}_V)$. The assumption of detailed balance \cite{hor3}
  reduces the possible terms in the Lagrangian, and it allows
for a quantum inheritance principle \cite{hor2}, since the
$(D+1)$-dimensional theory acquires the renormalization properties
of the $D$-dimensional one. Under the detailed balance condition
 the full action of Ho\v{r}ava-Lifshitz gravity is given by
\begin{eqnarray}
 S_g &=&  \int dt d^3x \sqrt{g} N \left\{
\frac{2}{\kappa^2}
(K_{ij}K^{ij} - \lambda K^2) \ \ \ \ \ \ \ \ \ \ \ \ \ \ \ \ \  \right. \nonumber \\
&+&\left.\frac{\kappa^2}{2 w^4} C_{ij}C^{ij}
 -\frac{\kappa^2 \mu}{2 w^2}
\frac{\epsilon^{ijk}}{\sqrt{g}} R_{il} \nabla_j R^l_k +
\frac{\kappa^2 \mu^2}{8} R_{ij} R^{ij}
     \right. \nonumber \\
&+&\left.    \frac{\kappa^2 \mu^2}{8( 3 \lambda-1)} \left[ \frac{1
- 4 \lambda}{4} R^2 + \Lambda  R - 3 \Lambda ^2 \right] \right\},
\label{acct}
\end{eqnarray}
where
\begin{eqnarray}
K_{ij} = \frac{1}{2N} \left( {\dot{g}_{ij}} - \nabla_i N_j -
\nabla_j N_i \right)
\end{eqnarray}
is the extrinsic curvature and
\begin{eqnarray} C^{ij} \, = \, \frac{\epsilon^{ijk}}{\sqrt{g}} \nabla_k
\bigl( R^j_i - \frac{1}{4} R \delta^j_i \bigr)
\end{eqnarray}
the Cotton tensor, and the covariant derivatives are defined with
respect to the spatial metric $g_{ij}$. $\epsilon^{ijk}$ is the
totally antisymmetric unit tensor, $\lambda$ is a dimensionless
constant and the variables $\kappa$, $w$ and $\mu$ are constants
with mass dimensions $-1$, $0$ and $1$, respectively. Finally, we
mention that in action (\ref{acct}) we have already performed the
usual analytic continuation of the parameters $\mu$ and $w$ of the
original version of Ho\v{r}ava-Lifshitz gravity, since such a
procedure is required in order to obtain a realistic cosmology
\cite{Lu:2009em,Minamitsuji:2009ii,Wang:2009rw,Park:2009zra}
(although it could fatally affect the gravitational theory
itself). Therefore, in the present work $\Lambda $ is a positive
constant, which as usual is related to the cosmological constant
in the IR limit.

Lastly, in order to incorporate the (dark plus baryonic) matter
component one adds a cosmological stress-energy tensor to the
gravitational field equations, by demanding to recover the usual
general relativity formulation in the low-energy limit
\cite{Sotiriou:2009bx,Chaichian:2010yi,Carloni:2009jc}. Thus, this
matter-tensor is a hydrodynamical approximation with its energy
density $\rho_M$ and pressure $p_M$ (or $\rho_M$ and its
equation-of-state parameter $w_M\equiv p_M/\rho_M$) as parameters.

Now, in order to focus on cosmological frameworks, we impose the
so called projectability condition \cite{Charmousis:2009tc} and
use a Friedmann-Robertson-Walker (FRW)  metric,
\begin{eqnarray}
N=1~,~~g_{ij}=a^2(t)\gamma_{ij}~,~~N^i=0~,
\end{eqnarray}
with
\begin{eqnarray}
\gamma_{ij}dx^idx^j=\frac{dr^2}{1- k r^2}+r^2d\Omega_2^2~,
\end{eqnarray}
where $ k=-1,0,+1$ corresponding  to open, flat, and closed
universe respectively. By varying $N$ and $g_{ij}$, we obtain the
equations of motion:
\begin{eqnarray}\label{Fr1fluid}
H^2 &=&
\frac{\kappa^2}{6(3\la-1)} \rho_M +\nonumber\\
&+&\frac{\kappa^2}{6(3\la-1)}\left[ \frac{3\kappa^2\mu^2
k^2}{8(3\lambda-1)a^4} +\frac{3\kappa^2\mu^2\Lambda
^2}{8(3\lambda-1)}
 \right]-\nonumber\\
 &-&\frac{\kappa^4\mu^2\Lambda  k}{8(3\lambda-1)^2a^2} \ ,
\end{eqnarray}
\begin{eqnarray}\label{Fr2fluid}
\dot{H}+\frac{3}{2}H^2 &=&
-\frac{\kappa^2}{4(3\la-1)} w_M\rho_M -\nonumber\\
&-&\frac{\kappa^2}{4(3\la-1)}\left[\frac{\kappa^2\mu^2
k^2}{8(3\lambda-1)a^4} -\frac{3\kappa^2\mu^2\Lambda
^2}{8(3\lambda-1)}
 \right]-\nonumber\\
 &-&\frac{\kappa^4\mu^2\Lambda  k}{16(3\lambda-1)^2a^2}\ ,
\end{eqnarray}
where we have defined the Hubble parameter as $H\equiv\frac{\dot
a}{a}$. The term proportional to $a^{-4}$ is the usual ``dark
radiation term'', present in Ho\v{r}ava-Lifshitz cosmology
\cite{Calcagni:2009ar,Kiritsis:2009sh}, while the constant term is
just the explicit cosmological constant. Finally, as usual,
$\rho_M$  follows the standard evolution equation
\begin{eqnarray}\label{rhodotfluid}
&&\dot{\rho}_M+3H(1+w_M)\rho_M=0.
\end{eqnarray}

As a last step, requiring these expressions to coincide with the
standard Friedmann equations, in units where $c=1$  we set
\cite{Calcagni:2009ar,Kiritsis:2009sh}:
\begin{eqnarray}
G_{\rm cosmo}&=&\frac{\kappa^2}{16\pi(3\lambda-1)}\nonumber\\
\frac{\kappa^4\mu^2\Lambda}{8(3\lambda-1)^2}&=&1,
\label{simpleconstants0}
\end{eqnarray}
where $G_{\rm cosmo}$ is the ``cosmological'' Newton's constant.
We mention that in theories with Lorentz invariance breaking (such
is Ho\v{r}ava-Lifshitz one) the ``gravitational'' Newton's
constant $G_{\rm grav}$, that is the one that is present in the
gravitational action, does not coincide with the ``cosmological''
Newton's constant $G_{\rm cosmo}$, that is the one that is present
in Friedmann equations, unless Lorentz invariance is restored
\cite{Carroll:2004ai}. For completeness we mention that in our
case
\begin{eqnarray}
G_{\rm grav}=\frac{\kappa^2}{32\pi}\label{Ggrav},
\end{eqnarray}
as it can be straightforwardly read from the action (\ref{acct}).
Thus, it becomes obvious that in the IR ($\lambda=1$), where
Lorentz invariance is restored, $G_{\rm cosmo}$ and $G_{\rm grav}$
coincide.

Using the above identifications, we can re-write the Friedmann
equations (\ref{Fr1fluid}),(\ref{Fr2fluid}) as
\begin{equation}
\label{Fr1b} H^2+\frac{k}{a^2} = \frac{8\pi G_{\rm
cosmo}}{3}\rho_M+ \frac{k^2}{2\Lambda a^4}+\frac{\Lambda}{2}
\end{equation}
\begin{equation}
\label{Fr2b} \dot{H}+\frac{3}{2}H^2+\frac{k}{2a^2} = - 4\pi G_{\rm
cosmo}w_M\rho_M- \frac{k^2}{4\Lambda a^4}+\frac{3\Lambda}{4}.
\end{equation}

\section{Generalized second law of thermodynamics}
\label{GSL}

Having presented the cosmological scenario of a universe governed
by Ho\v{r}ava-Lifshitz  gravity, we proceed to an investigation of
its thermodynamic properties, and in particular of the generalized
second thermodynamic law (although there is still missing a robust
proof for its general validity)
\cite{Unruh:1982ic,Unruh:1982ic1,Unruh:1982ic2,Unruh:1982ic3,Zhou:2007pz,
Wu:2008ir,Unruh:1982ic4,Saridakis:2009uu,Wall:2009wm,Unruh:1982ic5,Sadjadi:2010nu}.
As it is usual in the literature,  one considers the universe as a
thermodynamical system.  However, a priori it is not trivial what
should be the ``radius'' of the system in order to acquire a
consistent description. This subject becomes more important under
the light of use of black-hole physics \cite{Hawking} in a
cosmological framework \cite{Jacobson:1995ab,Jacobson:1995ab1},
that is connecting the `radius' and `area' of the universe with
its temperature and entropy respectively. As we mentioned in the
Introduction, the thermodynamic interpretation of field equations
can be applicable for any horizon, provided that the gravitational
theory is diffeomorphism invariant
\cite{Padmanabhan:2009vy,Padmanabhan:2009ry}, however the apparent
horizon is widely used in the literature either in flat
\cite{Frolov:2002va,Frolov:2002va1,Frolov:2002va2} or in non-flat
FRW geometry \cite{Cai:2005ra,Cai:2005ra1}. In the case of
Ho\v{r}ava-Lifshitz gravity, the breaking of diffeomorphism
invariance leaves us the apparent horizon as a reasonable choice.

The dynamical apparent horizon, a marginally trapped surface with
vanishing expansion, is in general determined by the relation
$h^{ij}\partial_i\tilde r\partial _j \tilde r=0 $, which implies
that the vector $\nabla \tilde r$ is null (or degenerate) on the
apparent horizon surface \cite{Bak:1999hd}. In a metric of the
form $
 ds^2=h_{ij}dx^idx^j+\tilde r^2d\Omega_2^2$, with $h_{ij}=\text{diag}(-1,a^2/(1-kr^2))$, $i,j=0,1
$, it writes \cite{Bak:1999hd}:
\begin{equation}
\label{apphor}
 \tilde{r}_A=\frac{1}{\sqrt{H^2+\frac{k}{a^2}}}.
\end{equation}
In summary, we consider the universe as a thermodynamical system
with the apparent horizon surface being its boundary.

Let us now proceed to the investigation of the generalized second
law of thermodynamics. We are going to examine whether the sum of
the entropy enclosed by the apparent horizon and the entropy of
the apparent horizon itself, is not a decreasing function of time.
Simple arguments suggest that after equilibrium establishes and
the universe background geometry becomes FRW, all the fluids in
the universe acquire the same temperature $T$
\cite{Setare:2007at}, which is moreover equal to the temperature
of the horizon $T_h$
\cite{Frolov:2002va,Frolov:2002va1,Frolov:2002va2,Cai:2005ra,Cai:2005ra1},
otherwise the energy flow would deform this geometry \cite{pa}.
However, although this will certainly be the situation at late
times, that is when the universe fluids and the horizon will have
interacted for a long time, it is ambiguous if it will be the case
at early or intermediate times (for instance the present CMB
temperature is of the order of 1 Kelvin, while the horizon
temperature is many orders of magnitude below this figure).
However, in order to avoid non-equilibrium thermodynamical
calculations, which would lead to lack of mathematical simplicity
and generality, the assumption of equilibrium, although
restricting, is widely disseminated in the generalized-second-law
literature
\cite{Padmanabhan:2009vy,Padmanabhan:2009ry,Frolov:2002va,Frolov:2002va1,Frolov:2002va2,Cai:2005ra,Cai:2005ra1,Setare:2007at,pa}.
Thus, we will follow this assumption and we will have in mind that
our results hold only at late times of the universe evolution.

In general, the apparent horizon $\tilde{r}_A$ is a function of
time. Thus, a change $d\tilde {r}_A$ in time $dt$ will lead to a
volume-change $dV$, while the energy and entropy will change by
$dE$ and $dS$ respectively. However, since in the two states there
is a common source $T_{\mu\nu}$, we can consider that the pressure
$P$ and the temperature $T$ remain the same
\cite{Frolov:2002va,Frolov:2002va1,Frolov:2002va2,Cai:2005ra,Cai:2005ra1}.
Such a consideration is standard in thermodynamics, where one
considers two equilibrium states differing infinitesimally in the
extensive variables like entropy, energy and volume, while having
the same values for the intensive variables like temperature and
pressure.
 In this case the first law of
thermodynamics writes $TdS=dE+PdV$, and therefore the  dark-matter
entropy reads (the universe contains only the dark-matter fluid
and we neglect the radiation sector.):
\begin{eqnarray}
\label{en2}dS_M&=&\frac{1}{T}\Big(P_M dV+dE_M\Big),
\end{eqnarray}
 where $V=4 \pi \tilde{r}_A^3/3$ is the volume of the system bounded by the
 apparent horizon and thus $ dV=4\pi\tilde{r}_A^2d\tilde
{r}_A$. We mention here that a thermodynamic identity of the form
of (\ref{en2}) has a universal validity, and the information about
a given system is only encoded in the form of the entropy
functional $S(E, V )$ \cite{Padmanabhan:2009vy}. In particular, in
the case of normal materials, this entropy arises because of our
coarse graining over microscopic degrees of freedom which are not
tracked in the dynamical evolution, however, in the case of
spacetime the existence of horizons for a particular class of
observers makes it mandatory that these observers integrate out
degrees of freedom hidden by the horizon.

 Dividing (\ref{en2}) by $dt$ we obtain
\begin{eqnarray}
\label{en3bb} \dot{S}_M&=&\frac{1}{T}\Big(P_M \,
4\pi\tilde{r}_A^2\dot{\tilde {r}}_A+\dot{E}_M\Big).
\end{eqnarray}
In this relation the time derivative of the apparent horizon
writes
\begin{equation}
\dot{\tilde r}_A=H\tilde{r}_A^3\Big[4\pi G_{\rm
cosmo}(1+w_M)\rho_M+\frac{k^2}{\Lambda a^4}\Big], \label{dotrh}
\end{equation}
as it easily arises differentiating the Friedmann equation
(\ref{Fr1b}) and using  (\ref{rhodotfluid}).

In order to connect the thermodynamically relevant quantities,
namely the energy $E_M$ and pressure $P_M$, with the
cosmologically relevant ones, namely the energy density $\rho_M$
and the pressure $p_M$, we can straightforwardly use:
\begin{eqnarray}
E_M&=&\frac{4\pi}{3}\tilde{r}_A^3\rho_M \label{Eitilde}\\
 P_M&=&w_M\rho_M.
  \label{Pilde}
  \label{Pilde}
\end{eqnarray}
 Inserting the
time-derivative of (\ref{Eitilde}), along with (\ref{Pilde}), into
(\ref{en3bb}), and using (\ref{rhodotfluid}), we obtain:
\begin{equation}
\dot{S}_M=\frac{1}{T}\left(1+w_{M}\right)\rho_{M}\,
4\pi\tilde{r}_A^2\left(\dot{\tilde {r}}_A-H\tilde{r}_A\right).
\label{sdotfluids2}
\end{equation}

At this stage, we have to connect the temperature of the matter
fluid $T$ to that of the horizon $T_h$. As we have said, although
at early and intermediate times these two temperatures do not
coincide in general, at late times, after the establishment of
equilibrium, they become equal, that is $T=T_h$. Now, $T_h$ has to
be related to the geometry of the universe. Note that although the
association of a temperature to a horizon was historically related
to black holes, it was soon realized that the study of quantum
field theory in any spacetime with a horizon shows that all
horizons possess temperatures
\cite{Fulling:1973md,Davies:1975th,PIdesitter}. In particular, an
observer who is accelerating through the vacuum state in flat
spacetime perceives a horizon and will attribute to it
\cite{Unruh:1976db} a temperature $T= \kappa_a / 2\pi$
proportional to her acceleration $\kappa_a$ (for a review, see
 \cite{Birrel:bkqft,mukhanovwini,Dewitt:1975ys,Takagi:1986kn,Sriramkumar:1999nw,Brout:1995rd,Wald:1999vt}).
We stress that the relation connecting the horizon temperature
with the geometry of the universe depends only on this geometry
and not on the gravitational sector of the scenario. Thus, for
spherical (FRW) geometry and according to the generalization of
black hole thermodynamics \cite{Hawking} to a cosmological
framework, the temperature of the horizon is related to its radius
through \cite{Padmanabhan:2009vy}
\begin{equation}
\label{Th}
 T_h=\frac{1}{2\pi\tilde{r}_A},
\end{equation}
either in general relativity
\cite{Jacobson:1995ab,Jacobson:1995ab1,Cai:2005ra,Cai:2005ra1}, in
modified gravitational theories \cite{Paranjape:2006ca}, or in
Ho\v{r}ava-Lifshitz gravity \cite{Cai:2009ph}.

As a last step, we have to connect the entropy of the horizon to
its radius $\tilde{r}_A$ (or equivalently to each area). As usual
this relation will be the corresponding one for black holes, but
with the apparent horizon instead of the black-hole horizon, and
thus it obviously depends on the particular gravitational sector
of the scenario. In the case of black holes in Ho\v{r}ava-Lifshitz
gravity, and under the detailed balance condition, this expression
is known \cite{Cai:2009qs,Cai:2009pe,Cai:2009ph} and thus its
cosmological application straightforwardly leads to:
 \begin{equation}
 \label{2eq13b}
 S_h =
 \frac{\kappa^2}{32\Lambda G_{\rm
cosmo}^2}\left[\Lambda \tilde
 r_A^2+2k\ln \left(\sqrt{\Lambda}\tilde
 r_A\right) \right],
 \end{equation}
 where we have also made use of the identifications
 (\ref{simpleconstants0}).
Note that in the extraction of this relation the authors have
neglected quantum buoyancy effects near the black hole horizon,
however such effects are expected to bring about lower-order
corrections  \cite{Bekenstein:1999bh}  and thus relation
(\ref{2eq13b}) captures the main behavior.

We mention that since we desire to remain general we have not
imposed the IR limit of Ho\v{r}ava-Lifshitz gravity, and thus in
the above relation we have not set $\lambda=1$. Although such a
choice would be very reasonable concerning the late-time epochs of
the universe, at very early times, where the universe is very
small and the UV features of the theory are revealed, the
divergence of $\lambda$ from 1 may be significant. However, since
the equilibrium assumption, which allowed us to equalize the
horizon temperature with that of the universe interior and perform
the above calculations, is justified only at late times, later on
we will impose the IR limit of the obtained relations. In other
words, investigating the intrinsic running character of $\lambda$
is very interesting, but it is not need to be performed in detail
for the present work, since in the end of the day one has to imply
the late time limit. Such an investigation would indeed be very
interesting as an independent work, without focusing on
thermodynamics, where one could impose Renormalization Group
methods to model its running.

As expected, in relation  (\ref{2eq13b}) the first term
corresponds to the standard (general relativity) result, while the
second term is the novel one, arising from Ho\v{r}ava-Lifshitz
gravity. Now, differentiating (\ref{2eq13b})  we finally acquire
\begin{equation}
\label{Sdoth}
 \dot{S}_{h}=\frac{\kappa^2}{16G_{\rm
cosmo}^2}\tilde{r}_A\dot{\tilde {r}}_A+\frac{\kappa^2 k}{16\Lambda
G_{\rm cosmo}^2 \tilde {r}_A}\dot{\tilde
 {r}}_A.
\end{equation}

Let us now proceed to the calculation of the total entropy
variation. Adding relations (\ref{sdotfluids2}) and (\ref{Sdoth}),
with $T$ given by (\ref{Th}), we find:
\begin{eqnarray}
\label{Sdottot}
 \dot{S}_{tot}\equiv \dot{S}_M+ \dot{S}_h=
8\pi^2\tilde{r}_A^3\left(\dot{\tilde
{r}}_A-H\tilde{r}_A\right)(1+w_M)\rho_M\nonumber\\
+\frac{\kappa^2}{16G_{\rm
cosmo}^2}\left(\tilde{r}_A+\frac{k}{\Lambda
\tilde{r}_A}\right)\dot{\tilde {r}}_A.
\end{eqnarray}
Thus, substituting also $\dot{\tilde {r}}_A$ by (\ref{dotrh}) we
result to:
\begin{widetext}
\begin{eqnarray}
\label{Sdottot2lam}
 \dot{S}_{tot}=
\tilde{r}_A^3H\left[8\pi^2\tilde{r}_A^3(1+w_M)\rho_M+\frac{\kappa^2
k}{16G_{\rm cosmo}^2\Lambda \tilde{r}_A} \right] \left[4\pi G_{\rm
cosmo}(1+w_M)\rho_M+\frac{ k^2}{\Lambda a^4}
\right]+\frac{\pi(3\lambda-1) \tilde{r}_A^4H k^2}{G_{\rm
cosmo}\Lambda a^4}\nonumber\\
+\left[\left(\frac{3\lambda-1}{2}\right)-1\right]
8\pi^2\tilde{r}_A^4 H(1+w_M)\rho_M.
\end{eqnarray}
\end{widetext}
Finally, at late times, where equilibration between the horizon
and the universe interior had been established, $\lambda$ has
taken its IR limit ($\lambda=1$) and thus the above relation
becomes:
\begin{widetext}
\begin{eqnarray}
\label{Sdottot2}
 \dot{S}_{tot}=
\tilde{r}_A^3H\left[8\pi^2\tilde{r}_A^3(1+w_M)\rho_M+\frac{2\pi
k}{G\Lambda \tilde{r}_A} \right] \left[4\pi G(1+w_M)\rho_M+\frac{
k^2}{\Lambda a^4} \right]+\frac{2\pi \tilde{r}_A^4H k^2}{G\Lambda
a^4},
\end{eqnarray}
\end{widetext}
where we have simplified the notation using $G$ for the Newton's
constant, since in the IR limit $G_{\rm cosmo}$ and $G_{\rm grav}$
coincide.

Relations (\ref{Sdottot2lam}) and (\ref{Sdottot2}) provide the
expression for the total entropy variation rate in a universe
governed by Ho\v{r}ava-Lifshitz gravity. Let us examine its sign.
Obviously, for a flat or closed universe $ \dot{S}_{tot}\geq0$ (we
remind that due to analytic continuation in the present work
$\Lambda$ is always positive and $\lambda\geq1$) and thus the
generalized second law of thermodynamics is valid. We mention that
this result holds independently of the matter equation-of-state
parameters and of the background geometry, provided it is FRW.

However, for an open universe ($k=-1$), in order to acquire $
\dot{S}_{tot}\geq0$ one has to have a non-zero matter component.
In the limiting case where matter is absent one can easily see
that $ \dot{S}_{tot}\geq0$ if $\Lambda\tilde{r}_A^2\geq1 $, or
using the explicit form for $\tilde{r}_A$ (from relations
(\ref{apphor}) and (\ref{Fr1b})) if $\Lambda a^2\geq1$. Thus, the
generalized second law is always violated for sufficiently small
scale factors (provided that we are still in the late-time
cosmological regime). The reason for this arises from the modified
horizon entropy relation (\ref{2eq13b}). Clearly, the presence of
the curvature as a coefficient of the correction term means that $
\dot{S}_{h}$ can be negative for $k=-1$ unless
$\Lambda\tilde{r}_A^2\geq1 $. Thus, since in the absence of matter
$ \dot{S}_{tot}$ and $ \dot{S}_{h}$ coincide, the aforementioned
violation of the generalized second law is implied.

The above analysis has been performed under the detailed-balance
condition, since in this case the relation for black-hole (and
thus for the apparent horizon) entropy, is well known. However,
detail balance is not at all a requirement and one can
straightforward go beyond it and repeat the aforementioned steps,
but with the new entropy relation that will arise from the
corresponding black hole solutions (and of course with the new
$\dot{\tilde r}_A$ that will arise from the new Friedmann
equations). Extracting the black hole entropy in
Ho\v{r}ava-Lifshitz gravity without detailed balance is an
independent task of its own, and one can have a variety of results
according to the specific form of detail-balance breaking
\cite{Kiritsis:2009rx}. Thus, one should in principle examine the
validity of the generalized second law in each case separately.
However, in all cases our result will be qualitatively maintained.
The reason is the following: As it has been extensively stated in
the literature \cite{Calcagni:2009ar,Kiritsis:2009sh},
Ho\v{r}ava-Lifshitz cosmology coincides completely with
$\Lambda$CDM if one ignores the curvature and imposes the IR
limit. Thus, all modifications in the black hole entropy of
various versions of non-detailed-balance Ho\v{r}ava-Lifshitz
gravity will depend on the curvature, while the entropy relations
will always include the standard (general relativity) basic term.
Note however that the sign of the correction term will also depend
on the sign of the coefficients of the higher curvature terms
needed to break the detail balance, and thus the overall sign will
be arbitrary \footnote{We thank S. Mukohyama for this comment.}.
Therefore, for all the specific versions of Ho\v{r}ava-Lifshitz
gravity, the resulting $ \dot{S}_{tot}$ will be different, but
always of the structure of (\ref{Sdottot2}), that is with a
conditionally negative contribution for all possible curvatures.

\section{Generalized second law of thermodynamics: an effective but inconsistent approach}
\label{GSLeff}

In the previous section we examined the validity of the
generalized second law of thermodynamics in Ho\v{r}ava-Lifshitz
cosmology, following the robust approach of considering a universe
containing only matter and incorporating the extra information of
Ho\v{r}ava-Lifshitz gravity through the modified entropy relation.
However, as we discussed in the Introduction, in the case of the
first law of thermodynamics some authors have chosen a different
approach, that is to absorb all the extra information of
Ho\v{r}ava-Lifshitz gravity into an effective dark energy fluid,
and thus considering a universe containing matter plus this
effective fluid, in a general relativity background
\cite{Wang:2009rw,Cao:2010xx}. However, this approach does not
possess the transparent theoretical justification of the previous
one, and can be misleading in the case where gravitational
phenomena are important (for example one could find wrong
black-hole properties in such an ``effective'' universe comparing
to those of the real one). In order to reveal this problematic
behavior and in order to be complete and comparable with part of
the literature, in this section we investigate the generalized
second law following this effective description of
Ho\v{r}ava-Lifshitz cosmology.

\subsection{Ho\v{r}ava-Lifshitz cosmology with effective dark energy fluid}

Observing the Friedmann equations
(\ref{Fr1fluid}),(\ref{Fr2fluid}) one can introduce an effective
dark-energy sector defining the energy density and pressure as
\begin{equation}\label{rhoDE0}
\rho_{DE}\equiv \frac{3\kappa^2\mu^2 k^2}{8(3\lambda-1)a^4}
+\frac{3\kappa^2\mu^2\Lambda ^2}{8(3\lambda-1)}
\end{equation}
\begin{equation}
\label{pDE0} p_{DE}\equiv \frac{\kappa^2\mu^2
k^2}{8(3\lambda-1)a^4} -\frac{3\kappa^2\mu^2\Lambda
^2}{8(3\lambda-1)},
\end{equation}
which after the identifications (\ref{simpleconstants0}) become
\begin{equation}\label{rhoDE}
\rho_{DE}\equiv \frac{1}{16\pi G_{\rm
cosmo}}\left(\frac{3k^2}{\Lambda a^4}+3\Lambda\right)
\end{equation}
\begin{equation}
\label{pDE} p_{DE}\equiv \frac{1}{16\pi G_{\rm
cosmo}}\left(\frac{k^2}{\Lambda a^4}-3\Lambda\right).
\end{equation}
 Therefore, the effective dark energy incorporates the
contributions from the dark radiation term (proportional to
$a^{-4}$) and from the cosmological constant. It is
straightforward to show that $\rho_{DE}$ satisfies the standard
evolution equation:
\begin{eqnarray}
\label{DEevol} \dot{\rho}_{DE}+3H(1+w_{DE})\rho_{DE}=0,
\end{eqnarray}
where $w_{DE}\equiv p_{DE}/\rho_{DE}$ is the effective dark energy
equation-of-state parameter. Finally, using the above definitions,
we can re-write the Friedmann equations (\ref{Fr1b}),(\ref{Fr2b})
in the standard form:
\begin{equation}
\label{Fr1c} H^2+\frac{k}{a^2} = \frac{8\pi G_{\rm
cosmo}}{3}(\rho_M+ \rho_{DE})
\end{equation}
\begin{equation}
\label{Fr2c} \dot{H}+\frac{3}{2}H^2+\frac{k}{2a^2} = - 4\pi G_{\rm
cosmo}( \rho_M+p_M+\rho_{DE}+p_{DE}).
\end{equation}

\subsection{Generalized second law of thermodynamics}

Following the effective approach to Ho\v{r}ava-Lifshitz cosmology
we assume that the universe contains the matter and the dark
energy fluid, in a general relativity background. Thus, all the
extra information is included inside $\rho_{DE}$ and $p_{DE}$, and
the problem is equivalent with the one in Einstein gravity but
with two fluids. Repeating the steps of section \ref{GSL} for two
fluids we easily find
\begin{eqnarray}
\dot{S}_{DE}&=&\frac{1}{T}\left(1+w_{DE}\right)\rho_{DE}\,
4\pi\tilde{r}_A^2\left(\dot{\tilde
{r}}_A-H\tilde{r}_A\right)\label{sdotfluids1c}\\
\dot{S}_M&=&\frac{1}{T}\left(1+w_{M}\right)\rho_{M}\,
4\pi\tilde{r}_A^2\left(\dot{\tilde
{r}}_A-H\tilde{r}_A\right)\label{sdotfluids2c},
\end{eqnarray}
while in this case differentiation of the Friedmann equation
(\ref{Fr1c}), using also (\ref{rhodotfluid}) and (\ref{DEevol}),
gives
\begin{equation}
\dot{\tilde r}_A=4\pi G_{\rm cosmo}H\tilde{r}_A^3\left[
(1+w_M)\rho_M+(1+w_{DE})\rho_{DE}\right]. \label{dotrhc}
\end{equation}
Note that as we discussed above, at late times, due to equilibrium
the temperature of the two fluids will be the same, and moreover
equal to the temperature of the horizon $T_h$. This horizon
temperature is still given by (\ref{Th}), since the corresponding
relation depends only on the cosmological geometry. Finally, since
the (effective) gravitational sector in now the standard general
relativity, the horizon entropy will be given by the corresponding
standard relation
 \begin{equation}
 \label{2eq13}
 S = \frac{4\pi \tilde r_A^2}{4G_{\rm grav}}.
 \end{equation}
Differentiating (\ref{2eq13}) we acquire
 \begin{equation} \label{Sdothc}
 \dot{S}_{h}=\frac{4\pi}{(3\lambda-1)G_{\rm cosmo}}\tilde{r}_A\dot{\tilde
 {r}}_A,
\end{equation}
 where we have also made use of the identifications
 (\ref{simpleconstants0}) and (\ref{Ggrav}).

 Adding relations
(\ref{sdotfluids1c}),(\ref{sdotfluids2c}) and (\ref{Sdothc}), we
find:
\begin{widetext}
\begin{eqnarray}
\label{Sdottot}
 \dot{S}_{tot}\equiv\dot{S}_{DE}+\dot{S}_M+ \dot{S}_h=
8\pi^2\tilde{r}_A^3\left(\dot{\tilde
{r}}_A-H\tilde{r}_A\right)\Big[(1+w_{DE})\rho_{DE}+(1+w_M)\rho_M\Big]+\frac{4\pi}{(3\lambda-1)G_{\rm
cosmo}}\tilde{r}_A\dot{\tilde
 {r}}_A,
\end{eqnarray}
\end{widetext}
and thus, substituting also $\dot{\tilde {r}}_A$ by (\ref{dotrhc})
we result to:
\begin{widetext}
\begin{eqnarray}
\label{Sdottot2clam} && \dot{S}_{tot}= 32\pi^3 G_{\rm cosmo}
\tilde{r}_A^6H\Big[(1+w_{DE})\rho_{DE}+(1+w_M)\rho_M\Big]^2 \ \ \ \ \ \ \ \ \ \ \ \
\ \ \ \ \ \ \  \ \ \ \ \ \ \ \ \ \ \ \ \ \ \ \ \ \ \  \ \ \ \ \ \ \ \ \ \ \ \ \ \ \ \ \ \ \  \ \ \ \ \ \ \ \ \ \ \ \ \ \ \ \ \ \ \ \nonumber\\
&& \ \ \ \ \ \ \ \ \ \ \ \ \ \ \ \ \ \ \  \ \ \ \ \ \ \ \ \ \ \ \
\ \ \ \ \ \ \  \ \ \ \ \ \ \ \ \ \ \ \ \ \, +
\left[\left(\frac{2}{3\lambda-1}\right)-1\right]8\pi^2H\tilde{r}_A^4\Big[(1+w_{DE})\rho_{DE}+(1+w_M)\rho_M\Big].
\label{heelp}
\end{eqnarray}
\end{widetext}
Finally, since the above results, based on the equilibrium
assumption, hold only for late times, we have to impose the IR
limit ($\lambda=1$) and therefore relation (\ref{heelp}) becomes
\begin{equation}
\label{Sdottot2c}
 \dot{S}_{tot}=
32\pi^3 G
\tilde{r}_A^6H\Big[(1+w_{DE})\rho_{DE}+(1+w_M)\rho_M\Big]^2\geq0,
\end{equation}
where we have simplified the notation using $G$ for the Newton's
constant, since in the IR limit $G_{\rm cosmo}$ and $G_{\rm grav}$
coincide.

We mention here that the effective approach can be
straightforwardly extended beyond the detail balance condition, in
a much more easier way than the approach of the previous section.
In particular, since one uses the standard (general relativity)
relation for horizon entropy, he does not need to examine the
black hole properties at all, but only to extract the Friedmann
equations and absorb all the new information in a modified
effective dark energy fluid. Thus, result (\ref{Sdottot2c}) will
be still valid, but $\rho_{DE}$ can now have many contributions
according to the specific extension beyond the detail balance.

A first observation is that $\dot{S}_{tot}$ is slightly different
that the one calculated in the previous section, and the two
results coincide only in the IR and zero-curvature limit. This
difference has been discussed in the case of the first
thermodynamical law \cite{Wei:2010ww}, where coincidence is
achieved in the same limit too. However, the qualitative
difference between the two approaches, that is the exact and the
effective one, is that now we obtain that $ \dot{S}_{tot}$ is
always non-negative in the IR limit and thus the generalized
second law would be always valid in the late-time universe. This
partially misleading result reveals that in subjects where gravity
is involved, one cannot follow a completely effective approach and
absorb all the gravitational phenomena inside conventional
components. Doing so one does not take into account the richness
of the effects of gravity. The fact that the results of the two
approaches coincide for zero curvature in the IR is
straightforwardly explained, since in this case the correction
term in the modified entropy relation (\ref{2eq13b}) is zero.

\section{Discussion and Conclusions}
\label{conclusions}

In this work we investigated the validity of the generalized
second law of thermodynamics in a universe governed by
Ho\v{r}ava-Lifshitz gravity. Considering the universe as a
thermodynamical system bounded by the apparent horizon, we
calculated the entropy time-variation of the universe-content as
well as that of the horizon, under the assumption of thermal
equilibrium which is expected to hold at late times. We stress
that we followed the theoretically robust approach, that is to
consider that the universe contains only matter, while the effect
of the novel gravitational sector of Ho\v{r}ava-Lifshitz gravity
was incorporated through the modified black-hole (and consequently
horizon) entropy. Additionally, although we performed our
calculations for an arbitrary $\lambda$, that is not only in the
IR limit of the theory, in the end, and in order to be consistent
with the equilibrium assumption, we have to focus on late times
and thus impose the IR limit. We found that at late times, under
detailed balance, for flat and closed universe the generalized
second law is always valid, while for open geometry it is only
conditionally valid. Going beyond detailed balance the generalized
second law is only conditionally valid for all curvatures.

The possible violation of the generalized second thermodynamical
law in Ho\v{r}ava-Lifshitz cosmology could lead to various
conclusions (apart from putting the generalized second law itself
or the equilibrium assumption into question). Although conditional
validity has been found for some horizon choices in other
cosmological scenarios too
\cite{MohseniSadjadi:2005ps,MohseniSadjadi:2005ps2}, one could
examine if using an alternative horizon would lead to a complete
validation. Usual alternative choices such is the Hubble radius
$H^{-1}$, or the future event horizon $R_h=\int^\infty_a
da/(Ha^2)$, lead to the same result, that is to conditional
validity. Nevertheless, one could still search for a suitable
horizon in order to acquire full validity. However, one should
have in mind that in Ho\v{r}ava-Lifshitz gravity the physical
meaning of a horizon might differ form that in general relativity.
In particular, in the later the static black-hole horizon and the
de Sitter horizon are Killing horizons, allowing for an
application of  Euclidean \cite{Hawking} or thermofield
\cite{Israel:1976ur} methods in order for the horizon temperature
to acquire a physical meaning. On the other hand, in
Ho\v{r}ava-Lifshitz gravity under the projectability condition,
there exists a preferred time slicing which is not orthogonal to
the horizon, which could furthermore be an emergent concept in the
IR limit \cite{Izumi:2009ry}, or particle-dependent (each particle
has its own ``light speed'' and thus it sees its own ``horizon'')
\cite{Kiritsis:2009rx}. Thus, the interpretation of the horizon
temperature may be different in the two theories. Finally, as was
discussed in \cite{Kiritsis:2009rx}, even without the
projectability condition, Ho\v{r}ava-Lifshitz gravity still shares
the uncertainties of non-relativistic theories, which can make the
notions of entropy and temperature not well-defined.

A second conclusion could be that the conditional violation of the
generalized second law is an additional problem along with the
discussed (not few) conceptual and theoretical problems of
Ho\v{r}ava-Lifshitz gravity
\cite{Charmousis:2009tc,Li:2009bg,Sotiriou:2009bx,Bogdanos:2009uj,
Koyama:2009hc,Papazoglou:2009fj}. However, the logarithmic
correction in the black-hole entropy, which is the cause of the
conditional violation, is not a so exotic term, and indeed it has
the same form with the quantum corrections on the standard result
calculated in loop quantum gravity (much earlier than the
appearance of Ho\v{r}ava-Lifshitz gravity)
\cite{Rov,Ash,Kau,Ghosh,Doma,Mei,Hod,Med,Sol,Kas,Das,Gour,Gour2,
Chat2,Cai:2008ys,Majhi:2008gi}. Thus, one could deduce that at
least this point could not be a problematic feature of
Ho\v{r}ava-Lifshitz gravity. Finally, one can still put into
question the fact that the black hole production inside the
apparent horizon and its effect on entropy, as well as possible
entropy bounds on matter, have been neglected.

In order to be comparable with part of the literature and for
completeness, we also performed the whole analysis following the
effective approach, that is to absorb all the extra information of
Ho\v{r}ava-Lifshitz gravity in an effective dark energy sector and
consider the resulting universe in a general relativity
background. Clearly this approach is not theoretically robust, and
the fact that our final result is partially different than the one
of the exact approach, offers an additional argument against the
naive effective incorporation of gravity in applications where
gravitational phenomena are important (such are the thermodynamic
properties of the universe). This was already known, and that is
why in all thermodynamic investigations in other modified
gravitational theories the authors followed the exact approach
instead of the effective one (see \cite{Paranjape:2006ca} and
references therein).

In conclusion, we see that the thermodynamic properties of
Ho\v{r}ava-Lifshitz gravity, either concerning the first law of
thermodynamic and its relation to the Friedmann equations, or
concerning the generalized second law, is a very interesting
subject that deserves further investigation. However, one has to
be careful and incorporate consistently the novel features of the
theory. In the end of the day, the knowledge acquired from these
studies will contribute to the acceptance or rejection of
Ho\v{r}ava-Lifshitz gravity as a description of nature.

\begin{acknowledgments}
The authors wish to thank Charalampos Bogdanos, Rong-Gen Cai,
Yungui Gong, Shinji Mukohyama,  Anzhong Wang, Shao-Wen Wei and an
anonymous referee, for useful discussions and valuable comments.
\end{acknowledgments}

\end{document}